\documentclass{epl}
\date{September 2000}
\shorttitle{Discrete energy landscapes and RSB at T=0}
\title{Discrete energy landscapes and replica symmetry breaking at zero temperature}
\author{F. Krzakala\footnote{krzakala@ipno.in2p3.fr} 
	\and O.C. Martin\footnote{martino@ipno.in2p3.fr} }
\institute{
  \inst { } Laboratoire de Physique Th\'eorique et Mod\`eles Statistiques,\\
Universit\'e Paris-Sud, b\^atiment 100, F-91405 Orsay,
France.
}


\begin{document}

\maketitle

\begin{abstract}
The order parameter $P(q)$ for disordered 
systems with degenerate ground-states is reconsidered. We propose
that entropy fluctuations lead to a trivial $P(q)$ at zero 
temperature as in the non-degenerate case, even if there are
zero-energy large-scale excitations (complex energy
landscape). Such a situation should arise
in the $3$-dimensional $\pm J$ Ising spin glass 
and in MAX-SAT. Also, we
argue that if the energy landscape is complex with a finite
number of ground-state families, then
replica symmetry breaking reappears at positive temperature.
\end{abstract}

\section{Introduction}
A number of disordered and frustrated systems~\cite{Young98} exhibit 
replica symmetry breaking (RSB)~\cite{MezardParisi87b}; physically,
this means that several macroscopically
different valleys contribute to the partition function in the thermodynamic
limit. Perhaps the most famous case is the Sherington-Kirkpatrick (SK)
model of spin glasses, but RSB is believed to arise in diluted
mean field models~\cite{VianaBray85,BanavarSherrington87,MezardParisi87c} too.
Furthermore, there is an ongoing debate
over the presence of RSB
in the Edwards-Anderson (EA) spin glass model in $d \ge 3$. 
Most Monte Carlo simulations~\cite{MarinariParisi99b}
give evidence in favor of RSB there, but it has been objected that
the temperatures used ($T \approx\, 0.7~T_c$) were too close to $T_c$.
Also, recent work going to 
zero temperature~\cite{Hartmann99b}
suggests that there is no RSB in the $d=3$ case, so this issue
remains open.

The order parameter for RSB is the presence of a non-trivial
probability distribution for 
the overlap $q$ between two configurations taken at
random according to their Boltzmann weight. In the mean field 
picture of {\it generic} random systems, there are macroscopically
different valleys whose
free-energies differ by $O(1)$, and thus $P(q)$ is non-trivial. 
However, since
the inter-valley part of $P(q)$ decreases linearly with the
temperature $T$ as $T \to 0$, the zero temperature limit
of $P(q)$ is trivial: only intra-valley overlaps survive. It
is often believed that this 
pattern is modified when the
energy is a discrete variable and the ground-states are highly
degenerate. Indeed, it is common lore that models with discrete
energies such as the $\pm J$ EA spin glass
or MAX-SAT near but below the transition point~\cite{BiroliMonasson00}
should have a non-trivial $P(q)$ if the mean field picture is
correct, whereas the presence of a trivial $P(q)$ is considered
to validate the droplet model~\cite{FisherHuse88}. Here we go against
this standard lore, and we will try to convince the reader that 
instead of examining $P(q)$ among ground-states it is more appropriate
to consider the energy landscape and determine whether there exist
macroscopically different valleys. We believe our conclusions apply 
not only to the $3$-dimensional $\pm J$ EA model, but also to other
discrete energy models such as
diluted fully-frustrated magnets, MAX-SAT, and 
diluted mean field $\pm J$ Ising spin glasses to name just a few. 

Our first claim is that $P(q)$ in these
models is trivial when considering ground-states 
or even finite energy excited states. Our second claim 
is that replica symmetry is broken
($P(q)$ is non-trivial) at arbitrarily low temperatures
in the thermodynamic limit. In effect, 
discrete and continuous energy models behave similarly
with respect to RSB. At the heart of our arguments is the
hypothesis that the effective number of ground-state
families does not grow with the system size. (Thus our
claims do not apply the $2$-dimensional spin
glasses.) The presentation that follows is given using the
$3$-dimensional EA model for definiteness, but the reader can easily
see how to rephrase the discussion for any 
model with a complex energy landscape.

\section{$P(q)$ for ground-states}

In the $3$-dimensional $\pm J$ EA Ising spin glass 
Hamiltonian,
\begin{equation}
H = - \sum_{<ij>} J_{ij} S_i S_j
\end{equation}
the $J_{ij}$ are quenched random variables, $J_{ij} = \pm 1$,
that couple nearest neighbor spins on a lattice of size 
$L \times L \times L$. The $J_{ij}$ are discrete so the
energy is also discrete. 
Furthermore, the local field felt by any spin can have
the values $\pm 6, \pm 4, \pm 2, 0$ only. A field value of $0$ 
leads to a spin that can flip without changing the energy. Since
a finite fraction of the spins on the lattice have this property,
there is a finite entropy per spin at the ground-state energy level: 
the number of different ground-states grows at least exponentially
with the lattice volume. 

A question of central importance is whether these
ground-states differ macroscopically. Unfortunately, this question
is ambiguous because of the
finite density of zero-energy droplets: if we flip all the spins 
in zero field, do we reach a macroscopically different ground-state?
We would like the answer to be {\it no}~. To make things simple,
suppose for this discussion that $L = \infty$; if all  
the ground-states differ only by zero-energy droplets 
containing say 1, 2, ..., $k$ spins, then we can say 
that they form a single
family of ground-states. In that case,
one expects $P(q)$ to be a delta function at
$q_{EA} < 1$. (Here and in all that follows, 
our statements are made modulo the up-down 
symmetry which flips all the spins.) If we let $k \to \infty$,
we still have a single family of ground-states
provided the frequency of zero-energy 
droplets decreases sufficiently rapidly as their size $k$ grows. This is
what happens in the ``hidden ferromagnet'' picture of spin
glasses due to Fisher and Huse~\cite{FisherHuse88}. In the same
vein, when the ground-states form {\it multiple} families,
one can hope that the ground-states belonging to the same
family will differ by finite size droplets only whereas
ground-states from different families should differ by at least
one infinite zero-energy droplet, plus any number
of finite size zero-energy droplets.

When $L$ is finite, it is even more difficult to define
ground-state families satisfactorily because
zero-energy droplets can arise with as large a $k$ as one
wants; there is thus inevitably some additional arbitrariness in one's 
definition of these families. One rather simple definition
was proposed and used by Hartmann~\cite{Hartmann99b}:
two ground-states are in the same family if it is possible
to go from one to the other by a sequence of zero-energy
droplets of size $k=1$. But this also means that two
ground-states differing by a zero-energy droplet of size $2$
may be put into two different families. Thus it seems
more appropriate to define families by clustering together
ground-states when their overlap is ``sufficiently large''.
In the case of one-step RSB, this is quite clean, the overlaps
being either small or large. When confronted with continuous
RSB though, the overlaps cannot naturally be split into small
and large values: there are families within families 
{\it ad infinitum}, and a
cut-off on the overlap has to be introduced, otherwise the number 
of families will necessarily be infinite. Because of this arbitrariness,
the clustering is less elegant than for one-step RSB;
nevertheless, this clustering-based procedure
can and has been implemented by Hed et al.~\cite{HedHartmann00}.
Their conclusion, as well as Hartmann's~\cite{Hartmann97},
who sampled $P(q)$ amongst ground-states, is that there
exists multiple macroscopically distinct families
with a finite (non-zero) probability as $L \to \infty$.

\begin{figure}
\resizebox{0.95\textwidth}{!}{
  \includegraphics{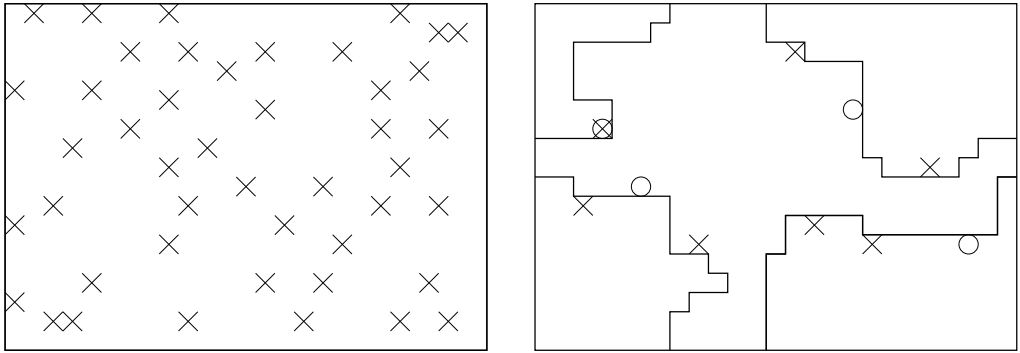}}    
\caption{a: All spins of ground-state $1$ having zero local field 
(crosses). b: We flip a system-size
connected cluster and reach ground-state $2$;
the solid line is the boundary of that cluster. Only those
zero local field spins that 
touch the boundary are relevant for the difference in entropies of the
two families of ground-states.}
\label{fig_flip}
\end{figure}

Given such a clustering-based definition of ground-state families,
we shall assume hereafter that the model under consideration can have
multiple families and thus there is
a finite probability of having macroscopically
different ground-states.
If, when taking the large $L$ limit, this
definition of families is good, one may for convenience 
qualitatively think
of a family as all those ground-states obtained from a given one by 
flipping zero-energy droplets of finite size.
On the contrary, when going from one
ground-state family to another, it is necessary to flip a
``system-size'' connected cluster of spins
that contains a finite fraction of the whole system. 
This is illustrated schematically
in Figure~\ref{fig_flip}.b where the solid lines
represent the boundary $B$ of such a system-size
cluster. (The picture is
for a cross-section of the $3$-dimensional lattice.)

In this kind of system, what does the distribution $P(q)$ of ground-state 
overlaps look like? For simplicity,
suppose that there are just two families,
containing ${\cal V}_1$ and ${\cal V}_2$ ground-states. 
For the purpose of comparing the ${\cal V}_i$, we consider
that each family is obtained from one reference
ground-state ${\cal C}_i$ by flipping any number of its zero-energy
droplets, taken to be non-adjacent (and thus non-interacting). 
Let ${\cal N}_i^{(k)}$ be the number of zero-energy
droplets of size $k$ of ${\cal C}_i$. 
Then we have
$Log_2 {\cal V}_i = {\cal N}_i^{(1)} + {\cal N}_i^{(2)} + \cdots $; clearly 
${\cal V}_i$, which is
crucial for the structure of $P(q)$, is
very sensitive to the exact number of zero energy droplets in
the family. We expect ${\cal N}_i^{(k)}$ to be self-averaging,
growing linearly with $L^3$. But there are also fluctuations 
in this number; if the central limit theorem holds, they will scale as
$L^{3/2}$, so that the total number of zero-energy droplets ${\cal N}_i$
of ${\cal C}_i$ has the form
\begin{equation}
\label{eq_fluct1}
{\cal N}_i = \sum_k {\cal N}_i^{(k)} = \alpha L^3 + x_i \sqrt{fL^3}
\end{equation}
where $x_i$ is a (Gaussian)
random variable. $P(q)$ has both inter and 
intra-family contributions. There are
${\cal V}_1^2 + {\cal V}_2^2$ ground-state pairs 
giving the intra-family part, and 
$2{\cal V}_1 {\cal V}_2$ pairs giving the inter-family part.
We then see that the intra-family part dominates 
the inter-family part by a factor
$2^{(x_1-x_2)\sqrt{f L^3}}$ 
(we took $x_1 > x_2$). Since this factor diverges
as $L \to \infty$, we find no RSB.

Upon closer examination, we see that it is necessary 
to be more careful because maybe
$x_1 \approx x_2$. In Figure~\ref{fig_flip}.a we show by crosses 
the spins of ground-state $1$ (${\cal C}_1$) having zero 
local field whose number is ${\cal N}_1^{(1)}$.
Figure~\ref{fig_flip}.b shows 
the boundary $B$ of the system-size
cluster that takes one from family $1$ to
family $2$. It is evident that except for the sites touching $B$,
the spins having zero local field are the {\it same} in family $1$ and 
in family $2$. These spins then do
not contribute to the difference ${\cal N}_1^{(1)} - {\cal N}_2^{(1)}$. 
However, for the spins that do touch $B$,
their local field can and usually will be different
in the two families. Denoting by $\cal A$ the area of $B$ (in fact
the number of spins touched by $B$), the argument that lead to
Eq.~\ref{eq_fluct1} suggests that ${\cal N}_1^{(1)} - {\cal N}_2^{(1)}$
scales instead as $\sqrt{{\cal A}}$. Extending this to all droplet
sizes, we obtain
\begin{equation}
\label{eq_fluct2}
{\cal N}_1 - {\cal N}_2 \approx y_{12} \sqrt{{\cal A}}
\end{equation}
where $y_{12}$ is another (Gaussian) random variable with finite variance.
We then see that the previous claim still holds, namely that
intra-family contributions dominate the inter-family ones
because ${\cal A}$ diverges when $L \to \infty$.
(By hypothesis, the number of spins that are flipped when
going from one family to another scales as
$L^3$, so ${\cal A}$ grows at least as $L^2$.) 

This presentation was given for two ground-state
families. Extending it to any finite number of families is
straight-forward; what really matters for the argument
to go through is for 
${\cal N}_i - {\cal N}_j$ to diverge for any pair of families.
This will happen 
as long as the number of families
does not grow too quickly with $L$.
To stay as simple as possible, we will continue to assume
that the model has a finite number of 
ground-state families. Then
the infinite volume limit of
$P(q)$ is trivial (delta functions at $\pm q_{EA}$)
because of entropy fluctuations amongst the different families. 
But the trivial nature of 
$P(q)$ should not be considered as evidence against
having finite-energy large-scale excitations ($\theta = 0$ in
the language of the droplet model),
nor against the mean field picture for that matter,
as we shall soon see. Finally, there is every 
reason to expect the reasoning to apply to all discrete
models, on lattices or not. For instance in the $\pm J$
Viana-Bray model, $\cal A$ scales as the total number of spins, 
so the triviality of $P(q)$ for 
ground-states should be easier to see than in the EA case.

In a numerical study, the factor 
$2^{y_{12}\sqrt{\cal A}}$ should be large. Then for most settings
of the disorder variables $J_{ij}$, the inter-family peak in 
the distribution $P_J(q)$ of ground-state overlaps 
will be small compared to the intra-family term. However, 
$y_{12}$ has a finite probability density at $0$, so with probability
$1/\sqrt{\cal A}$ the two peaks will be of comparable size.
Thus the $L$ dependence of 
$P(0)$ ($P_J(0)$ averaged over disorder)
should go as $1/\sqrt{\cal A}$ which is faster than
$1/L$. Hartmann~\cite{Hartmann99b} indeed finds that
$P(0)$ decays, and his fits indicate a $1/L^{1.25}$ dependence.
Taken at face value within our framework, this means that
${\cal A}(L) \approx L^{2.5}$.

In the picture we have reached,
$P(q)$ is trivial but the energy landscape is ``complex''; by that
we mean that macroscopically different ground-states appear
(taking into account of course the global up-down symmetry).
The numerical results~\cite{Hartmann97,Hartmann99b,HedHartmann00}
in support of this picture no longer seem mysterious from
a mean field perspective, and lead us to conclude
that models with discrete and continuous energies behave
similarly: neither have RSB at $T=0$. Note that RSB at $T=0$ implies
a complex energy landscape, but the reverse is not true.

\section{$P(q)$ for excited states}
The arguments we gave go beyond just ground-states.
Consider the energy landscape of the system, and think of it
as composed of different valleys. In this
section, we focus on those valleys whose bottom energies
are less or equal to
some arbitrary fixed cut-off $E_{max}$.
We can consider $P(q)$ for overlaps
either among the bottoms of these valleys, or among
{\it all} spin configurations of energy
$E \le E_{max}$. (Note that the configurations
are {\it not} weighted by a Boltzmann factor.) 
Interestingly, the result is the 
same in both cases, so let us begin with the first one which is simpler.

When $E_{max}$ is equal to the ground-state energy, the
valley bottoms considered are precisely the set of ground-states, and
we have clustered these into families. Since defining these families
required some arbitrary choices, we see that the same
must be true for defining valleys. Thus let valleys
be also defined by clustering according to overlaps.
The we generalize to valleys a property we used before for ground-state
families: for each discrete energy level $E_i \le E_{max}$, 
we assume that there is
a finite probability of having multiple valleys whose
bottom energy is $E_i$.
Furthermore, as before, 
we suppose that the number of such
valleys remains finite as $L \to \infty$. Then, regardless
of their energy, the
valley bottoms play the same role as the ground-state
families of the previous section. In consequence,
when comparing any two valley bottoms,
one will dominate the other. The situation is thus identical to the one
we had when looking only at ground-states, and we conclude that
$P(q)$ for overlaps between the valley bottoms is trivial
as $L \to \infty$. 

Two comments are in order.
First, which valley dominates is a random process.
As these valleys have energies differing only by
$O(1)$, they should have the same statistical properties, so
the probability of dominating should be uniform among the different
valleys. But since
we also expect the typical number of valleys to increase
with energy, the winning valley will 
most likely be near the cut-off $E_{max}$.
The second comment is that 
since entropy fluctuations lead to single valley
dominance, the reasoning suggests
that $P(q)$ is trivial at positive temperatures
also. This extrapolation will turn out to be too na\"\i ve:
having a positive temperature requires taking $E_{max}$ to $\infty$,
and then we have to deal with an infinite number of valleys. Before
looking into that case,
let us go on and investigate as promised
the effet of positive energy excitations within the valleys.

Thus, we now consider {\it all} the configurations 
satisfying $E \le E_{max}$. We can estimate the
size of the valleys (bottoms and higher energy configurations) 
by including the possibility of 
flipping positive energy droplets. Because the energy is discrete
and bounded by $E_{max}$,
for each valley we can only flip a {\it finite} number of such 
droplets. If we consider all the configurations obtained
from a valley bottom by flipping $k$ such droplets, 
we find that their number is $O(L^{3k})$ times larger than
the number at the bottom of the valley. Obviously this power-law
factor cannot beat the
$2^{y_{ij}\sqrt{\cal A}}$ factor, and
positive energy droplets lead to ``negligible'' corrections
to the valley entropies. The fundamental reason is
that the cut-off $E_{max}$ puts a bound on the number $k$ of positive
energy droplets that can be flipped. 

\section{$P(q)$ at a positive temperature}
When $T>0$, the number of positive
energy droplets
that can be simultaneously excited grows linearly 
with the volume of the lattice. Since
we have no reason to excite a finite size droplet of energy $E$ rather than
a system-size cluster of the same energy, we see that we have
to let $E_{max}$ go to $\infty$. Then we have a infinite number
of valleys to consider! To make progress, 
we must describe how these valleys are distributed and then
estimate their contributions to the partition function.

As was done implicitly when discussing the excited
states in valleys, we heuristically view a valley as
a ground-state plus all of its possible
(finite size) droplet excitations. This is
certainly only part of the picture even when just considering
the ground-state energy level, but it will do for our presentation.
We begin with a statistical description of the number
of valleys whose bottoms are at the energy level $E$. In our complex
energy landscape, there are more and more such valley bottoms
as $E$ grows; we denote by $\rho_E(E)$ their density.
In the ``random energy model''~\cite{Derrida81}
and in the SK model, this density grows exponentially with $E$; 
here we need not be so explicit, we just take
\begin{equation}
\label{eq_rho_E}
{\rho}_E(E) = e^{S(E)} = e^{S(E^*) + (E-E^*)S'(E^*) + \cdots }
\end{equation}
where $S(E)$ is a smoothly growing function and
$E^*$ is any large argument. The important hypothesis
we make is that $S(E)$
does not grow with $L$; this must be the case
if there is to be a finite number of valleys whose bottoms are 
at the different (discrete) values of $E$.

Now we want to determine the density $\rho_F$ 
of free-energies of these valleys when the temperature
is $T$. First, the free-energy $F_i(T)$ of valley $i$ 
is defined via:
\begin{equation}
\label{eq_F_Z}
e^{-\frac{F_i(T)}{T}} = e^{ -\frac{E_i}{T}} 
\sum_{\cal C}{'} e^{-\frac{E(C)-E_i}{T}}
\end{equation}
where $\sum_{\cal C}^{'}$ denotes the sum over all configurations
$\cal C$ belonging to valley $i$, and
$E_i$ is the energy of the valley's bottom. 
Second, we make the hypothesis that the free-energy $F_i(T)$ of valley $i$,
when measured with respect to the
energy $E_i$ at its bottom, is a random variable behaving
for instance according to what the central limit theorem would
predict were it applicable:
\begin{equation}
\label{eq_F}
F_i(T) = E_i + L^3 f(T) + x_i \sqrt{L^{3}}
\end{equation}
Here $f(T)$ is the (self-averaging) free-energy density,
and $x_i$ is a (Gaussian) random variable whose variance depends
on $T$. This ansatz is based on the idea that the valleys 
are statistically similar, and so the second factor on the
right-hand-side of Eq.~\ref{eq_F_Z} has no
statistical dependence on the value of $E_i$. Put simplistically,
there is no way to know within one of these valleys whether or not
it contains the ground-state.

Given our ansatz, the density ${\rho}_F(F)$
of the valley free-energies can now be computed. From Eq.~\ref{eq_F},
${\rho}_F(F)$ is obtained by a convolution of ${\rho}_E(E)$ with a Gaussian
distribution $G(x)$:
\begin{equation}
\label{eq_rho_F}
{\rho}_F(F)= \int_{-\infty}^{\infty} {\rho}_E(E)dE
\int_{-\infty}^{\infty} G(x) ~ 
{\delta}(F- E - L^3f(T) - x \sqrt{L^3} ) dx
\end{equation}
First, use the $\delta$ function to perform the integral over $x$.
Then, because
$\rho_E(E)$ grows steeply, the integral's main contribution
comes from the neighborhood of $E^*$, a large value
of $E$. Finally, we use Eq.~\ref{eq_rho_E};
defining $1/T^* = S'(E^*)$, the integrand becomes
a Gaussian in $E$. Performing the integral then leads to
\begin{equation}
{\rho}(F)= e^{\frac{F-F^*(T)}{T^*}}
\end{equation}
where $F^*(T) \approx L^3 f(T)$ is a reference free-energy. The important
point is that $\rho_F(F)$ grows exponentially with $F$: we obtain
exactly the form given by the random energy model and by mean field theory.
For such a $\rho_F$, the lowest (valley) free-energies differ by
$O(1)$, and so multiple valleys contribute to the partition
function even as $L \to \infty$. The conclusion is that 
for positive temperatures below $T^*$, there is RSB, and $P(q)$
is non-trivial.

After all this is said and done, we realize that we have made the
same mistake here as when we estimated $P(q)$ for ground-states.
Indeed, in Eq.~\ref{eq_F}, the $x_i$
of the different valleys are {\it correlated}. What is relevant
is the scaling of the {\it differences} in the valley
free-energies. Generalizing the argument used for ground-states,
we replace Eq.~\ref{eq_F} by
\begin{equation}
F_i(T) - F_j(T) = E_i - E_j + y_{ij}\sqrt{{\cal A}_{ij}}
\end{equation}
Taking any scaling form for ${\cal A}_{ij}$, we substitute
in Eq.~\ref{eq_rho_F}
the term $x \sqrt{L^3}$ by a term $y \sqrt{{\cal A}(L)}$; up to 
unimportant prefactors, this leads to the same exponential law
for $\rho_F$ as before, and thus to RSB.

It is also possible to extend the calculation using a very
different ansatz. Assume that energy differences between valley
bottoms are not $O(1)$ but instead scale as $L^{\theta}$.
Then one finds that the free-energy differences also scale
as $L^{\theta}$.

\section{Summary and conclusions}
In a general system with quenched disorder, we say that its
energy landscape is ``complex'' if there are 
macroscopically different valleys with $O(1)$ energy differences.
When the energy is discrete, (as in the 
$\pm J$ Ising spin glasses), the ground-states can be highly degenerate,
but we expect to be able to clusterize them into families.
If the number of such familiess remains finite in the
infinite volume limit, then we expect just one family
to dominate all the others by its size. This effect is simply due
to entropy fluctuations among the different families,
and leads to the absence of replica symmetry breaking
at zero temperature. The absence of RSB (that is a trivial
overlap distribution $P(q)$) at zero temperature does
not necessarily validate the droplet model. Instead, our
over-all message can be summarized as follows:
``A system having valleys with $O(1)$ {\it energy} differences 
(a complex energy landscape)
should lead to a free-energy landscape with $O(1)$ {\it free-energy} 
differences, and thus to RSB''. It is thus not relevant to look at
$P(q)$ at zero temperature, nor even $P(q)$ among 
finite energy excited states; the energy landscape
seems to be a better object to consider
when trying to distinguish the droplet and
mean-field pictures. Amuzingly, we see that
the discrete nature of the energies plays no role in determining
whether there is RSB or not, be-it at zero or at
finite temperature.

Our point of view allows one to understand a number of
numerical results on the $\pm J$ EA model~\cite{Hartmann97,Hartmann99b}. 
Furthermore, it suggests that when going to very low 
temperatures~\cite{KatzgraberPalassini00}, it is important
to control the limit $L \to \infty$ before taking $T \to 0$.
Finally, we predict
that a careful study of ground-states in MAX-SAT
will give a trivial $P(q)$ rather than a one-step RSB pattern.

\acknowledgments
We thank H. Hilhorst and E. Marinari for
discussions and J.-P. Bouchaud for his detailed comments. F.K 
acknowledges financial support from the MENRT, 
and O.C.M. support from the Institut Universitaire de
France. The LPTMS is an Unit\'e de Recherche Mixte de
l'Universit\'e Paris~XI associ\'ee au CNRS.

\bibliographystyle{prsty}
\bibliography{../../../Bib/references}

\end{document}